\documentclass[aps,prl,showpacs,twocolumn,amsmath]{revtex4}
\usepackage{graphicx}
\newcommand{\bq}{\begin{equation}}
\newcommand{\eq}{\end{equation}}
\newcommand{\bn}{\begin{eqnarray}}
\newcommand{\en}{\end{eqnarray}}
\newcommand{\bsub}{\begin{subequations}}
\newcommand{\esub}{\end{subequations}}

\usepackage{ifpdf}
\usepackage{bm}
\usepackage{latexsym}
\usepackage{color}
\usepackage{pstricks}
\usepackage{figsize}

\usepackage{ifthen}
\usepackage{ifpdf}

\usepackage{latexsym}
\usepackage{amssymb}
\usepackage{bm}

\ifpdf
\usepackage{graphicx}
\usepackage{epstopdf}
\else
\usepackage{graphicx}
\usepackage{epsfig}
\fi

\begin{document}

\newcommand{\abs}[1]{\left|#1\right|}
\newcommand{\sinc}{\mbox{sinc}}
\newcommand{\const}{\mbox{const}}
\newcommand{\trc}{\mbox{trace}}
\newcommand{\intt}{\int\!\!\!\!\int }
\newcommand{\ointt}{\int\!\!\!\!\int\!\!\!\!\!\circ\ }
\newcommand{\ar}{\mathsf r}
\newcommand{\im}{\mbox{Im}}
\newcommand{\re}{\mbox{Re}}
\newcommand{\e}{\epsilon}
\newcommand{\T}{\mathcal{T}}

\newcommand{\eexp}{\mbox{e}^}
\newcommand{\bra}{\left\langle}
\newcommand{\ket}{\right\rangle}

\newcommand{\mass}{\mathsf{m}}
\newcommand{\Mass}{\mathsf{M}}

\newcommand{\tbox}[1]{\mbox{\tiny #1}}
\newcommand{\bmsf}[1]{\bm{\mathsf{#1}}}
\newcommand{\amatrix}[1]{\matrix{#1}}
\newcommand{\pd}[2]{\frac{\partial #1}{\partial #2}}

\newcommand{\ee}{\end{eqnarray}}
\newcommand{\be}[1]{\begin{eqnarray}  {\label{#1}}}
\newcommand{\hide}[1]{}
\newcommand{\drawline}{\begin{picture}(500,1)\line(1,0){500}\end{picture}}
\newcommand{\bitem}{$\bullet$ \ \ \ }
\newcommand{\Cn}[1]{\begin{center} #1 \end{center}}
\newcommand{\mpg}[2][1.0\hsize]{\begin{minipage}[b]{#1}{#2}\end{minipage}}
\newcommand{\mpgt}[2][1.0\hsize]{\begin{minipage}[t]{#1}{#2}\end{minipage}}
\newcommand{\putgraph}[2][0.30\hsize]{\includegraphics[width=#1]{#2}\vspace{0.3cm}}
\newcommand{\avg}[1]{\left\langle #1 \right\rangle}
\newcommand{\rd}[1]{\textcolor{red}{#1}}

\title{ Theory of the Quantum Hall Insulator}

\author{Roi Levy}
\author{Yigal Meir}

\affiliation{Department of Physics, Ben Gurion University, Beer Sheva 84105, ISRAEL}


\begin{abstract}
{The quantum Hall transition  \cite{vonklizing80} is one of the simplest and most studied quantum phase transitions. Nevertheless, the experimental observation of a new phase in this regime, the quantum Hall insulator,  still remains a puzzle since the first report more than a decade ago
\cite{Shahar1,Shahar2,Hilke98,Hilke99,Hilke99a,visser06,delang07},
as it is in contradiction with all theoretical studies based on microscopically coherent quantum calculations \cite{Entin-Wohlman,Pryadko,Zulicke,cain}.
In this work we introduce into the coherent quantum theory a new ingredient -- rare incoherent events, in a controlled manner. Using both direct numerical solutions and real-space renormalization, we demonstrate that these decoherence events stabilize the elusive quantum Hall insulator phase,  which becomes even more stable with increasing temperature and voltage bias, in agreement with experiments.
}
\end{abstract}
\date{\today}
 \maketitle
The integer quantum Hall effect \cite{vonklizing80} has been a paradigm for two-dimensional quantum phase transitions: a transition between the quantum Hall phase, characterized by a quantized Hall resistance $\rho_{xy}$ and a vanishing longitudinal resistance $\rho_{xx}$, and an insulator,
characterized by diverging $\rho_{xx}$ and $\rho_{xy}$.
  This transition can be intuitively understood within the semiclassical description of the quantum Hall effect, valid at strong magnetic fields. In such fields electrons follow equipotential lines, which due to disorder are localized around valleys of the potential, at low energies, and peaks, at high energies. Near the potential saddle points, where such trajectories get close to each other (Fig.~\ref{fig:1}a), electrons can tunnel from one trajectory to another, where the tunneling probability depends on the characteristics of the saddle point \cite{Fertig87}. The critical energy is the energy at which there will be a trajectory that percolates through the system. Thus the quantum Hall transition may be described as a quantum percolation transition \cite{Trugman,Milnikov,Dubi}. Such a network model \cite{Chalker} of saddle points has also formed the basis for extensive numerical calculations and for real-space renormalization group (RSRG) calculations \cite{Galstyan97}. These calculations, consistent with other numerical studies of the quantum Hall transition \cite{Huckenstein} and with field theoretical renormalization group analysis \cite{Pruisken} indeed demonstrate the existence of  a critical value ${\cal T}_c$ of  the average transmission through a saddle point ${\cal T}$,  separating two stable phases - the quantum Hall phase (${\cal T}>{\cal T}_c$) and the insulating phase (${\cal T}<{\cal T}_c$).

However, the experimental discovery of a new phase, the quantum Hall insulator\cite{Shahar1,Shahar2,Hilke98,Hilke99,Hilke99a,visser06,delang07}, characterized by a quantized  $\rho_{xy}$ but an exponentially large $\rho_{xx}$ (inset of Fig.~\ref{fig:4}a), has  eluded theoretical understanding. Microscopically coherent quantum mechanical calculations, applied to this regime, such as
numerical simulations \cite{Pryadko} or RSRG analysis \cite{Zulicke}, have demonstrated that there is no quantum Hall insulator phase
 in this system, consistent with other studies \cite{Entin-Wohlman,cain}, except perhaps for small enough systems \cite{sheng}. Taking dephasing into account phenomenologically, by addressing a finite-size system, mimicking the finite dephasing length,  Pryadko and Auerbach \cite{Pryadko} have concluded that $\rho_{xy}$ of the entire system will be given by its  value for that length scale, which, in principle, could be exponentially large - again in contradiction with the experimental observation.

\begin{figure}[]
\centering
\includegraphics[clip,width=\hsize ]{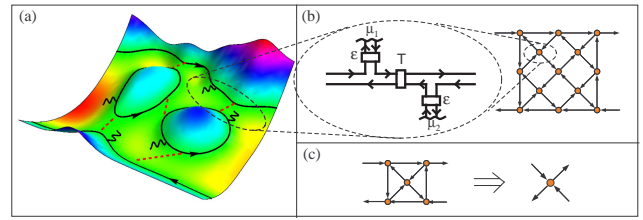}
 \caption{(color online) \textbf{(a)} Electron trajectories in the potential landscape. In strong magnetic field electrons follow equipotential lines (thick lines), and may undergo incoherent scattering events (wiggly lines). Near saddle points an electron can tunnel from one equipotential line to another (broken curves), with a corresponding tunneling probability $\T$. Such a junction, including the possibility of quantum tunneling and of incoherent scattering,  is represented by the elementary unit \textbf{(b)}, characterized by $\mathcal{T}$ and the decoherence parameter $\epsilon$. These elementary units, represented by orange dots, are connected in a network (here of size $3\times3$).  \textbf{(c)} The renormalization group procedure:  mapping five elementary units connected in Whetstone bridge form, representing the system in (a), onto a single effective elementary unit.}
\label{fig:1}
\end{figure}

Here we incorporate into the quantum mechanical calculation rare incoherent scattering events the electron undergoes during its motion along the equipotential lines (wiggly lines in Fig.~\ref{fig:1}a). These  events are introduced into the model via  phase randomizing, current conserving reservoirs \cite{buttiker} (see Fig.~\ref{fig:1}b), which means that for every electron that enters such a reservoir there is an electron that leaves it. However, the phase of the outgoing electron has no correlation with the phase of the incoming electron, so that once an electron enters such a reservoir, interference effects are destroyed. The basic unit in our network model consists of a saddle point (described by a scattering matrix, with random phase due to disorder) straddled  by two current-conserving reservoirs,  with probability $\epsilon$ to enter each reservoir (Fig.~\ref{fig:1}b). Accordingly, the transport parameters of the system can be calculated via a standard quantum multi-terminal scattering approach \cite{Landauer-Buttiker}, where two of the terminals represent the incoming and outgoing current, two -- the voltage probes which measure the Hall voltage, and the rest -- the current-conserving reservoirs, leading to decoherence. For $\epsilon=0$ this model reduces to the previously studied coherent network model \cite{Chalker}.
The coherent transmission from left to right, $T_{RL}$, is thus the quantum probability that the electron traverse the system in this direction, without entering any of the phase randomizing reservoirs. $T_{LR}$, $R_L$ (the reflection from left to left) and $R_R$ are similarly defined.
Consequently, the  probability $\epsilon$ to enter the reservoir determines the amount of decoherence in the system, characterized, e.g. by the decoherence length $L_\phi$, the typical length an electron travels before losing memory of its original phase. When $\epsilon=0$, no electron enters the phase randomizing reservoirs, and
the system is fully coherent, $L_\phi\rightarrow\infty$. When it is unity the system is classical,  with no coherent transport.
Tuning this probability from zero to unity allows us to probe the crossover between the fully coherent to the fully incoherent regimes.
\begin{figure}[]
\centering
\includegraphics[clip,width=0.44\hsize ]{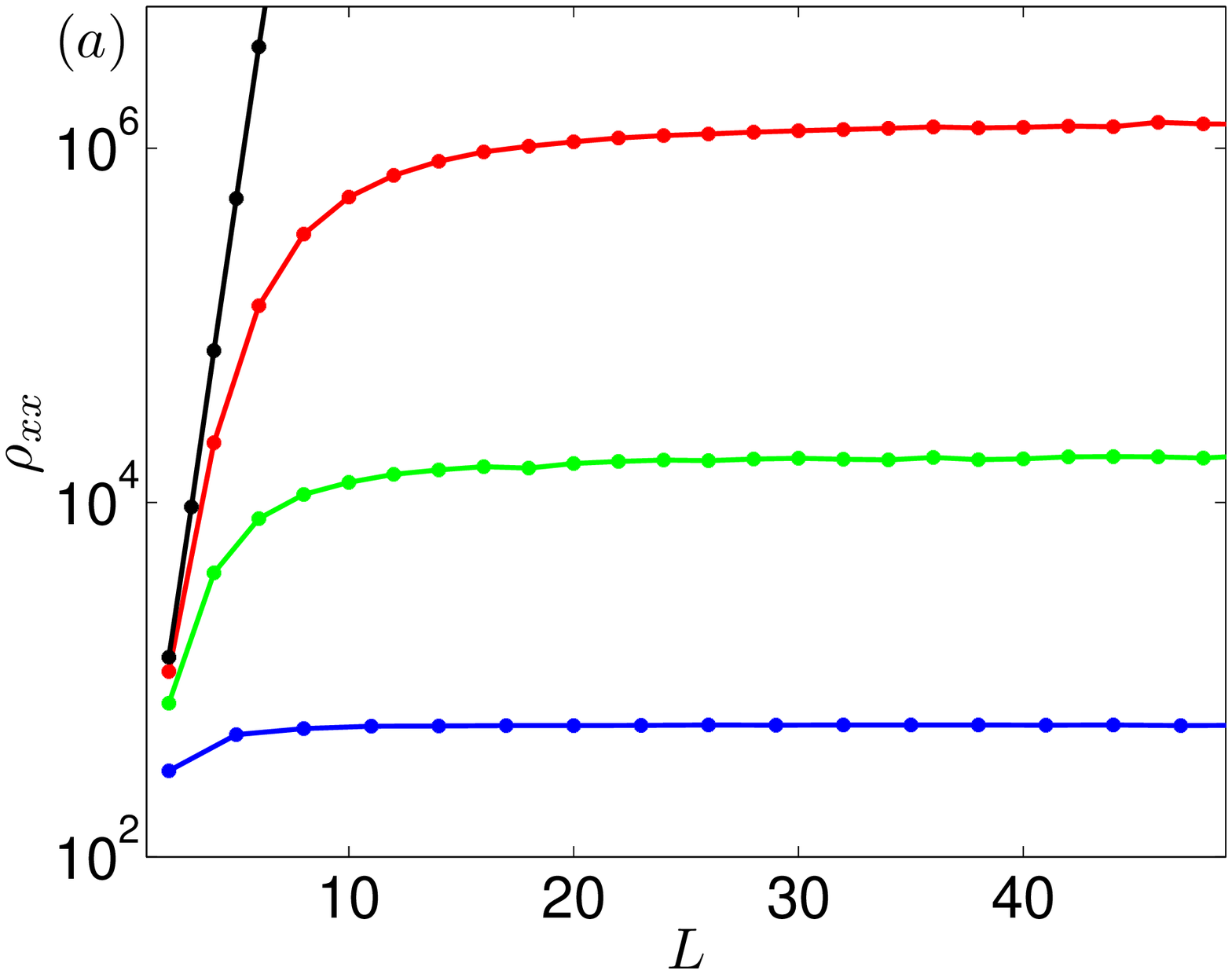}
\includegraphics[clip,width=0.44\hsize ]{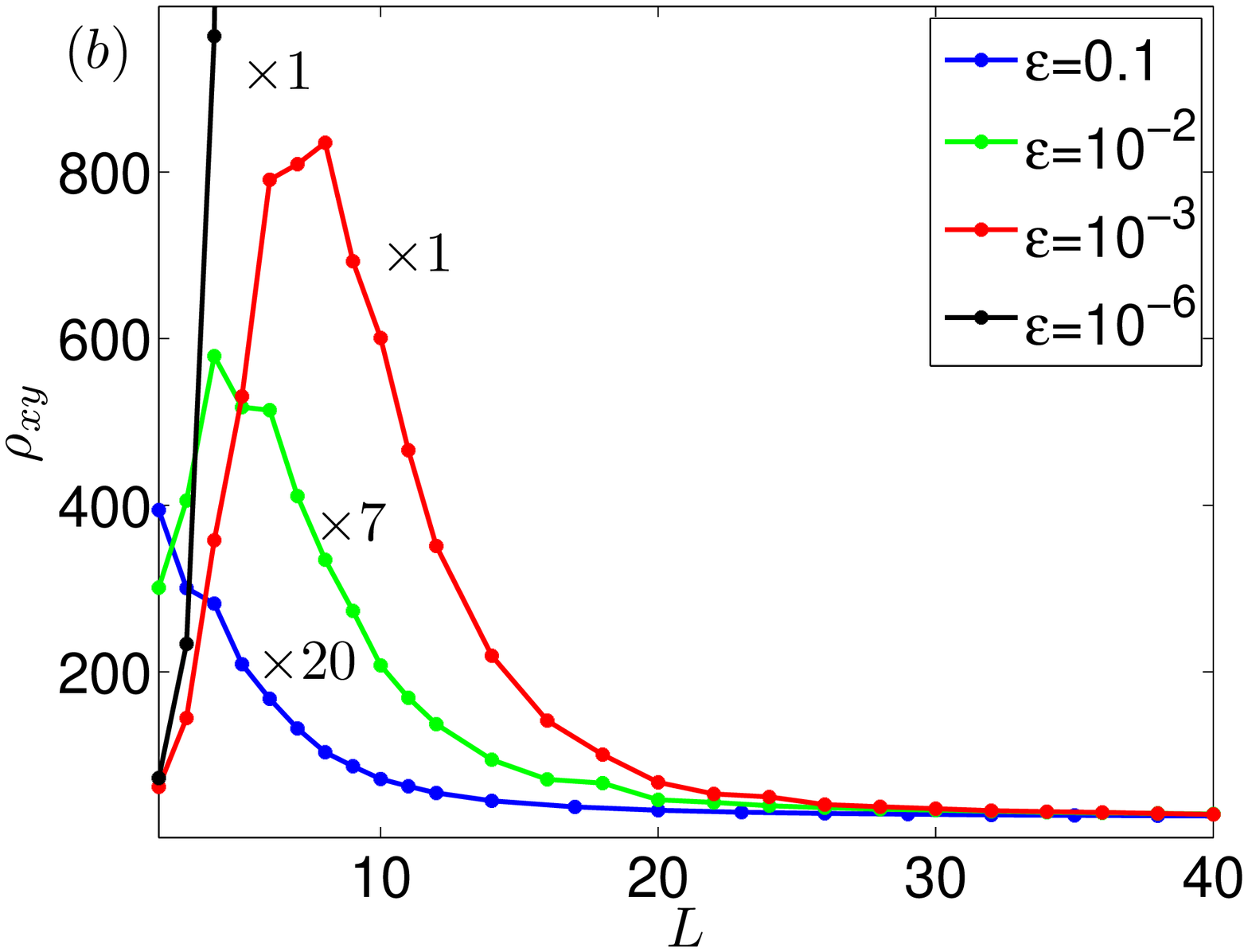}
\includegraphics[clip,width=0.44\hsize ]{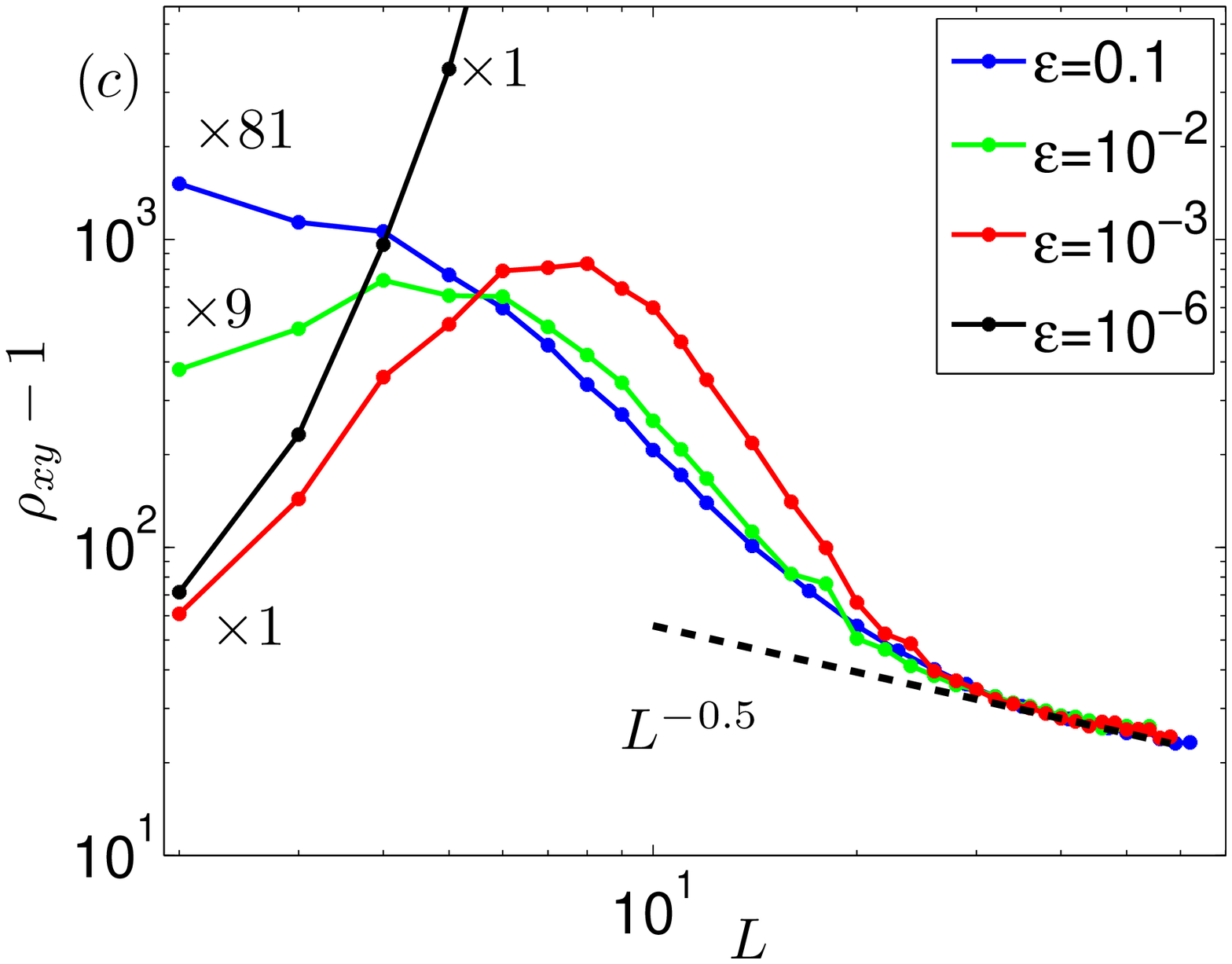}
\includegraphics[clip,width=0.44\hsize ]{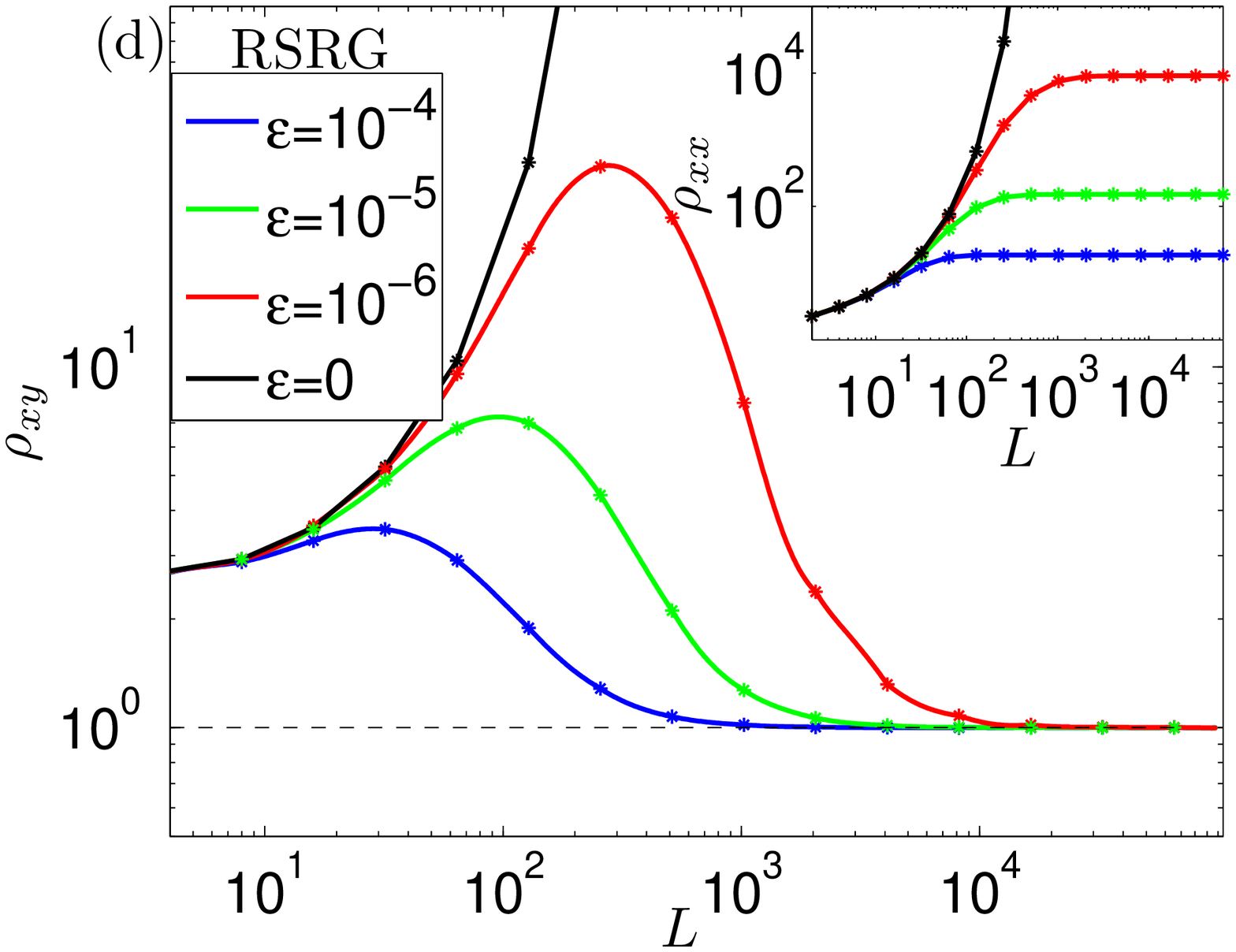}
\caption{(color online) $\rho_{xx}$ \textbf{(a)} and $\rho_{xy}$ \textbf{(b)} from the network model as a function of network size for several values of the dephasing parameter $\epsilon$ (values depicted in (b)), and average transmission of a single scatterer $<\T>=0.13$. While $\rho_{xx}$ saturates at $L\gg L_\phi$, $\rho_{xy}$ exhibits a nonmonotonic behavior. \textbf{(c)} Same date as in (b),  plotted as $\rho_{xy}-1$, multiplied by a constant, as a function of size on a double-log scale. At large sizes all curves collapse onto a single curve, confirming that for large sizes $\rho_{xy}\rightarrow1$. \textbf{(d)} Results of the real-space renormalization method:  $\rho_{xy}$ (inset $\rho_{xx}$) as a function of system size for different values of the initial dephasing parameter $\epsilon$ , with the same initial transmission $<\T>=0.4$, exhibiting the same behavior of both $\rho_{xy}$ and $\rho_{xx}$ as the network model. In the limit $L\gg L_\phi$,  $\rho_{xy}$ is quantized to unity, while $\rho_{xx}$ can be arbitrarily large. This is the quantum Hall insulator phase. }
\label{fig:2}
\end{figure}

We first solve numerically for $\rho_{xx}$ and $\rho_{xy}$ for a network of size $L\times L$ (Fig.~\ref{fig:1}b),  for different values of the average transmission though a saddle point,  {$\cal T$} and the decoherence parameter $\e$. $\rho_{xx}\equiv (1-T)/T$ is determined by the effective transmission $T$ through the whole system, while $\rho_{xy}$, as in the experiment, is determined by $V_H$, the anti-symmetric component of the difference between the chemical potentials at the upper and lower branches of the structure with respect to magnetic field, which is nonzero due to the chiral nature of the problem: $\rho_{xy}\equiv V_H/I$, where $I$ is the current. Because both $\rho_{xx}$ and $\rho_{xy}$ are exponentially distributed, we have used a logarithmic
average \cite{Zulicke} to calculate the effective renormalized values.
 The values of the saddle-point transmission probabilities {$\cal T$ are taken from a wide distribution, with a predefined average,  while the coupling to the current conserving reservoirs $\epsilon$ are taken from a delta distribution.

In Fig.~\ref{fig:2} we plot $\rho_{xx}$ (a) and $\rho_{xy}$ (b) as a function of the size of the system, for different values of the dephasing parameter $\e$. In the insulating phase (${\cal T}<{\cal T}_c$), and in the absence of decoherence events ($\e\rightarrow0$), both $\rho_{xx}$ and $\rho_{xy}$ increase exponentially, consistent with previous studies \cite{Pryadko}. In the presence of decoherence,  $\rho_{xx}$ first increases with system size (for $L<L_\phi$), and then saturates, as one expects for a classical system. Surprisingly, while $\rho_{xy}$ also initially increases, for $L<L_\phi$, it reaches a maximum and then decreases. For the samples with larger $\e$ (smaller $L_\phi$) $\rho_{xy}$ decreases all the way to unity (all resistance values are expressed in units of $h/e^2$, where $h$ is the Planck constant and $e$ the electron charge). While for samples with smaller $\e$, $\rho_{xy}$ has not yet reached the asymptotic regime, $L\gg L_\phi$, we demonstrate in Fig.~\ref{fig:2}c the collapse of all the curves, when plotted as  $\rho_{xy}-1$ vs $L$. This indeed confirms that, independent of $\e$ (or $L_\phi$), $\rho_{xy}$ scales as $\rho_{xy}=1+c_\e f\left(L/L_\phi\right)$, with $f(x)\sim1/\sqrt{x}$ for large $x$. This phase, where $\rho_{xy}$ is quantized to unity, and $\rho_{xx}$ could be exponentially large, is the elusive quantum Hall insulator phase.

In order to address asymptotically large systems $L\gg L_\phi$, we employ the RSRG method. In this approach one replaces a part of the system, containing several saddle points, by one effective saddle point, whose characteristics depend on those of the saddle points included in that subsystem (Fig.~\ref{fig:1}c). By following the dependence of the effective saddle-point transmission probability, ${\cal T}$, upon renormalization (see below), one can demonstrate that the coherent system flows either toward the insulating state (${\cal T}=0$) or towards the quantum Hall state (${\cal T}=1$) \cite{Zulicke}, an observation consistent with field theoretical renormalization group analysis \cite{Pruisken} and numerical calculations \cite{Huckenstein}.
In the presence of decoherence,  both ${\cal T}$ and $\epsilon$ are renormalized, as the probability for an incoherent scattering event increases as the system size increases. The basic cell for the RSRG method used here is  a five-unit cell in a Wheatstone bridge setup (representing the system depicted in Fig.~\ref{fig:1}a), which is
transformed into a single unit with effective transmission $\mathcal{T}_{eff}$ and decoherence  parameter $\epsilon_{eff}$. $\epsilon_{eff}$ is defined in terms of all the coherent parts of the scattering matrix,
\be{1drg}
\epsilon_{eff}=1-\frac{T_{LR}+T_{RL}+R_R+R_L}{2}.
\ee
Our RSRG transformation thus consists of two recursion equations, one for the total current (transmission) and
one for the decoherence parameter $\epsilon$. We solved these equations
using Monte-Carlo sampling rejection method \cite{MC}, and averaged over $2.5*10^5$ realizations.
Starting from some initial distributions of $\mathcal{T}$ and  $\epsilon$, this step is repeated to generate the next generation distributions, which are used as input to the next iteration. Fig.~\ref{fig:2}d depicts  the resulting $\rho_{xx}$ and $\rho_{xy}$ as a function of the system size $L$, for a given value of  $\T$, and for several values of $L_\phi$ (or $\epsilon$) on the insulating side. Similarly to the direct numerical solution, $\rho_{xx}$ saturates for $L>L_\phi$, and $\rho_{xy}$ displays the same intriguing behavior: initially it grows exponentially, but then around $L\simeq L_\phi$ it exhibits a maximum and then decreases back to unity.

\begin{figure}[]
\centering
\includegraphics[clip,width=0.49\hsize ]{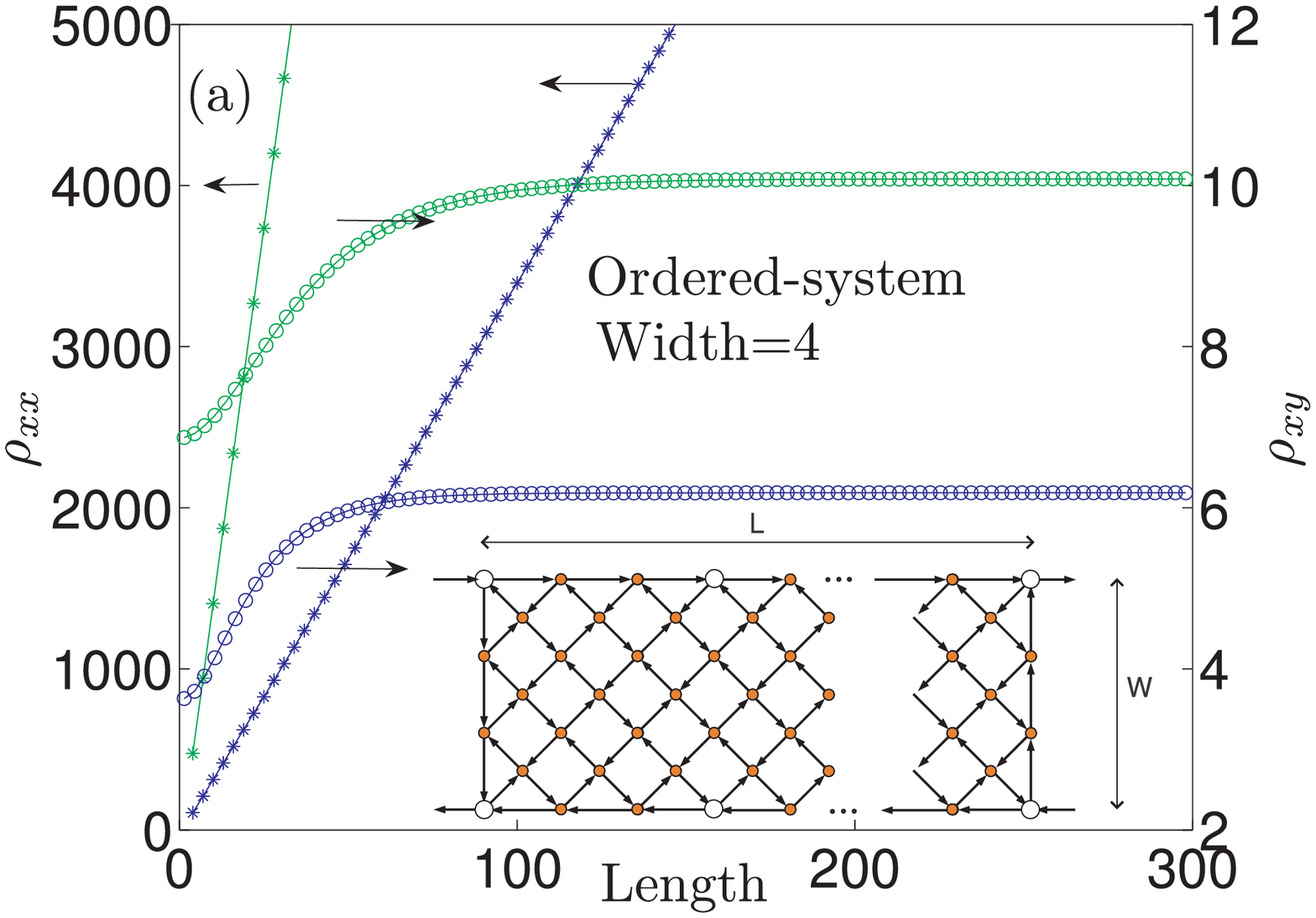}
\includegraphics[clip,width=0.46\hsize ]{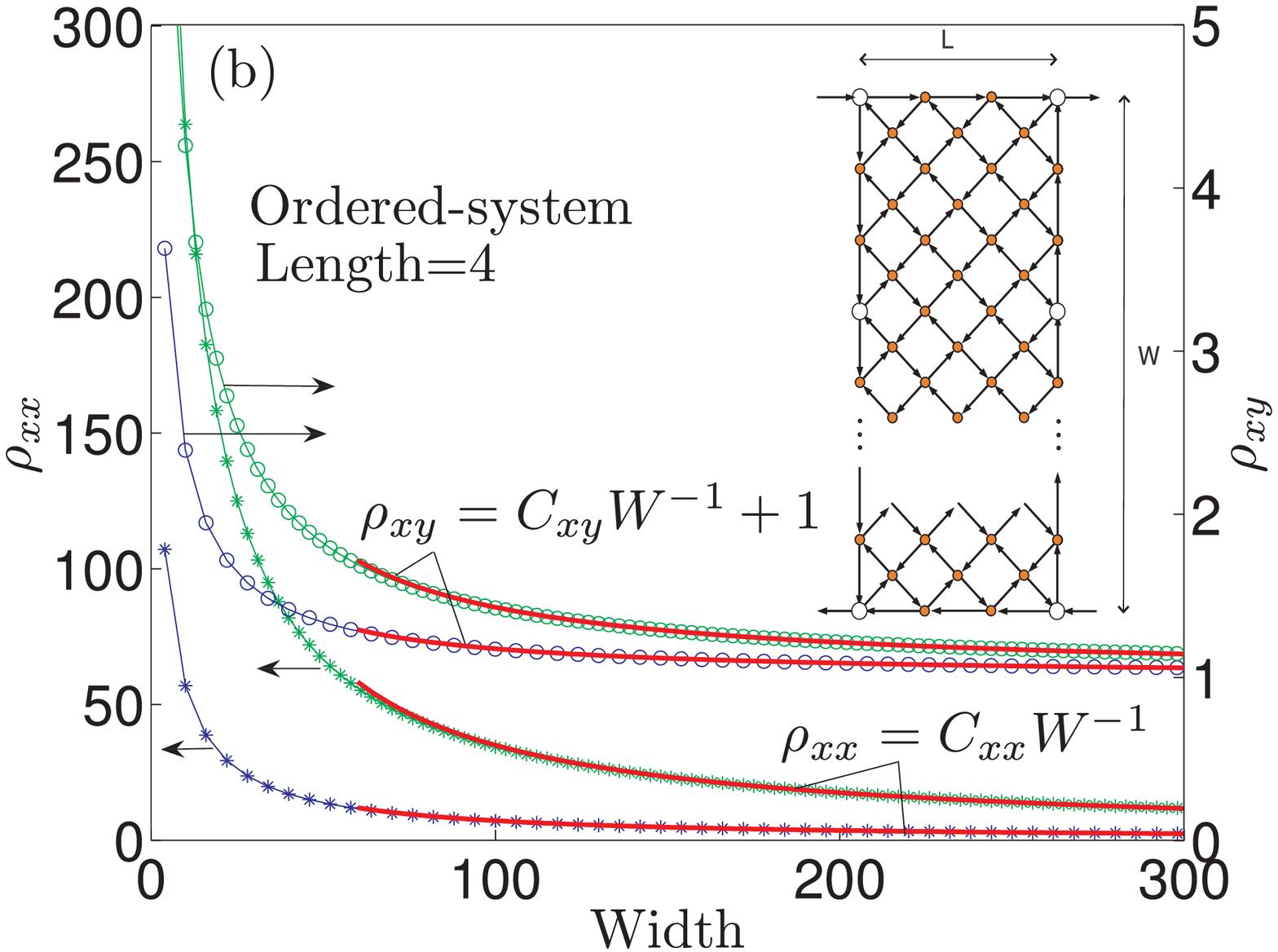}
\caption{(color online)  \textbf{(a)} $\rho_{xy}$ and $\rho_{xx}$ for ordered system (same phase and transmission value for all the scatterers) as a function of length for system of fixed width $W=4$, the green color is for $\T=0.05$  and the blue is for $\T=0.1$. Here dephasing occurs only at the corners of the elementary squares ($4\times4$ networks, plotted as empty dots) for which $\epsilon=1$.  $\rho_{xx}$ grows linearly with the length while $\rho_{xy}$ saturates. \textbf{(b)} For a system of fixed length, as the system becomes wider $\rho_{xy}$ goes to unity ($C_{xy}$ equals $43$ and $17.6$ for $\T=0.05$ and $\T=0.1$ respectively). $\rho_{xx}$ goes to zero as expected, with $C_{xx}$ equals $3500$ and $718$ for $\T=0.05$ and $\T=0.1$, respectively.}
\label{fig:3}
\end{figure}

While the decrease of $\rho_{xy}$ towards unity, as the system size increases, seems apriori surprising, one can show that it is a direct consequence of the rules of connecting resistors in series and parallel. To demonstrate this point we have calculated $R_{xx}$ and $R_{xy}$ (the longitudinal and Hall resistances for non-square systems) for a stack of coherent ordered squares, each of size $L_0\times L_0$, where the only decoherence scatterers are at the corners of these elementary squares (see inset of Fig.~\ref{fig:3}), connected either in series (Fig.~\ref{fig:3}a) or in parallel  (Fig.~\ref{fig:3}b). For the series connection  we find that $R_{xx}$ increases linearly with the system length $L$, as expected, and $R_{xy}$ saturates, while for the parallel connection, both  $R_{xx}$ and $R_{xy}$ decreases as $1/W$, where $W$ is the width of the sample, the former towards zero, again as expected, while the latter towards unity. This same behavior is also observed for rectangular disordered network-model systems (not shown). Both these behaviors of $R_{xy}$ can be readily understood. We first note that since $I=T V$, where $V$ is the voltage difference between source and drain, and $R_{xx}=(1-T)/T$, one can write $R_{xy}=(V_H/V)(R_{xx}+1)$. For the series connection, when $L\gg L_\phi$, one can think of the system as consisting of $L/L_\phi$ coherent segments, connected incoherently. Thus the voltage drop on each segment is $VL_\phi/L$. Because the Hall voltage of each segment is linearly dependent on the voltage drop across that segment, it scales like $1/L$, and as $R_{xx}$ grows linearly with $L$, $R_{xy}$ remains constant. On the other hand as the system width $W$ increases, for constant length, $R_{xx}$ decreases as $1/W$, and thus the above relation dictates that $R_{xy}$ also decreases (since $V_H$ is bound from above by $V$). In fact, in this limit, the upper chemical potential becomes dominated by the source chemical potential, while the lower chemical is dominated by the drain chemical potential, and thus $V_H/V$ approaches unity as the width increases. Consequently, as $R_{xx}$ approaches zero with increasing width, $R_{xy}$ approaches unity.
 (This is in contrast with the analysis of Ref.~\onlinecite{Pryadko}, which claims that  $R_{xy}$ remains constant as the width increase, while $R_{xx}$ decreases towards zero, violating the above relation between $R_{xx}$ and $R_{xy}$.)
 $\rho_{xy}$ for large square of size $L\times L$ can then be obtained by first making the system longer, of length $L$, such that its $R_{xy}$ does not change any more, and then increasing its width to $L$, so that the Hall resistance decreases towards unity. This is a generalization of the case considered in Ref.~\onlinecite{Shimshoni97}, which effectively connected in series and parallel puddles of $R_{xy}=1$, leading to $\rho_{xy}=1$ for the full system. This observation is also consistent with the two-phase approach \cite{Dikhne}.

\begin{figure}[]
\centering
\includegraphics[clip,width=0.46\hsize ]{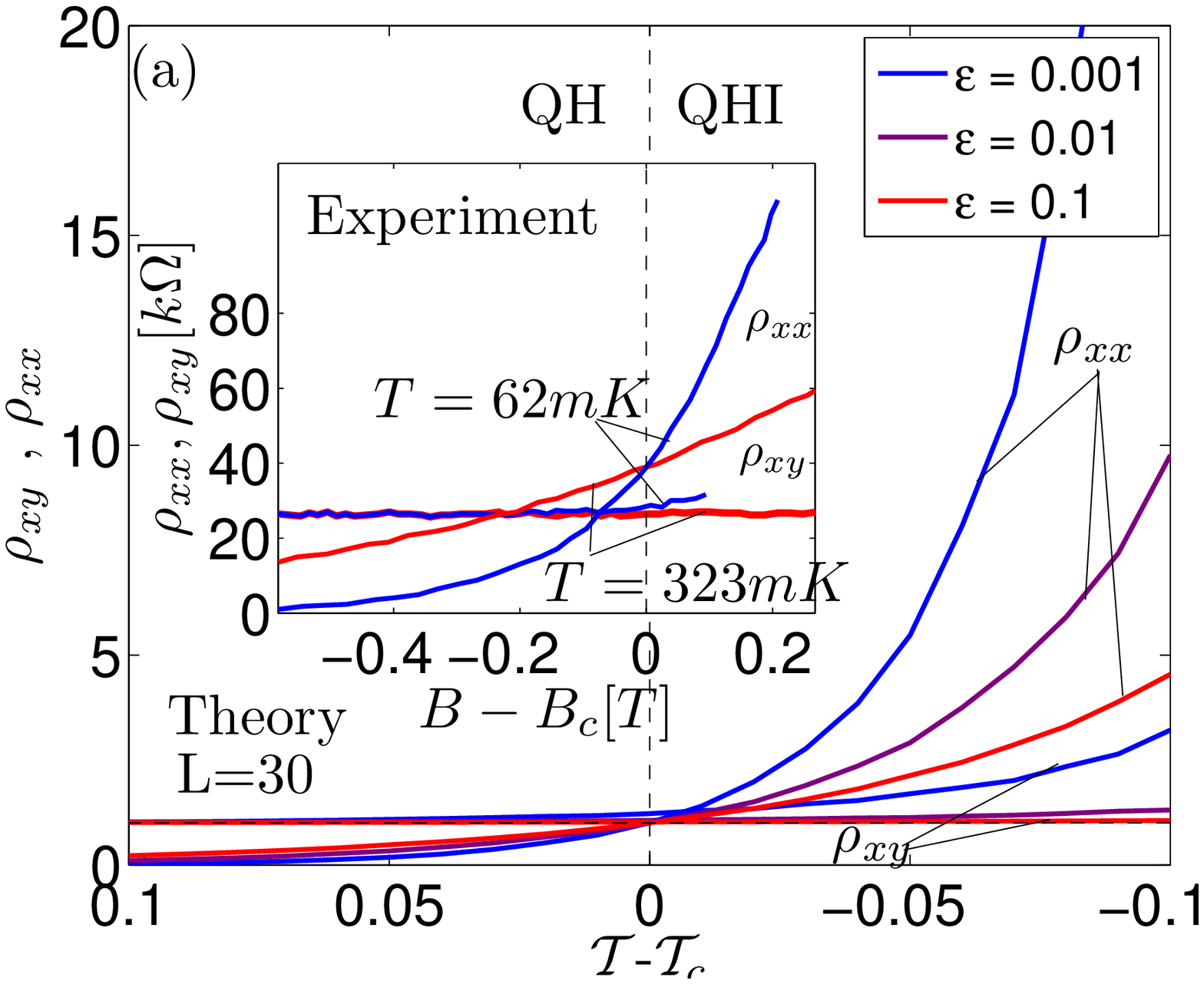}
\includegraphics[clip,width=0.45\hsize ]{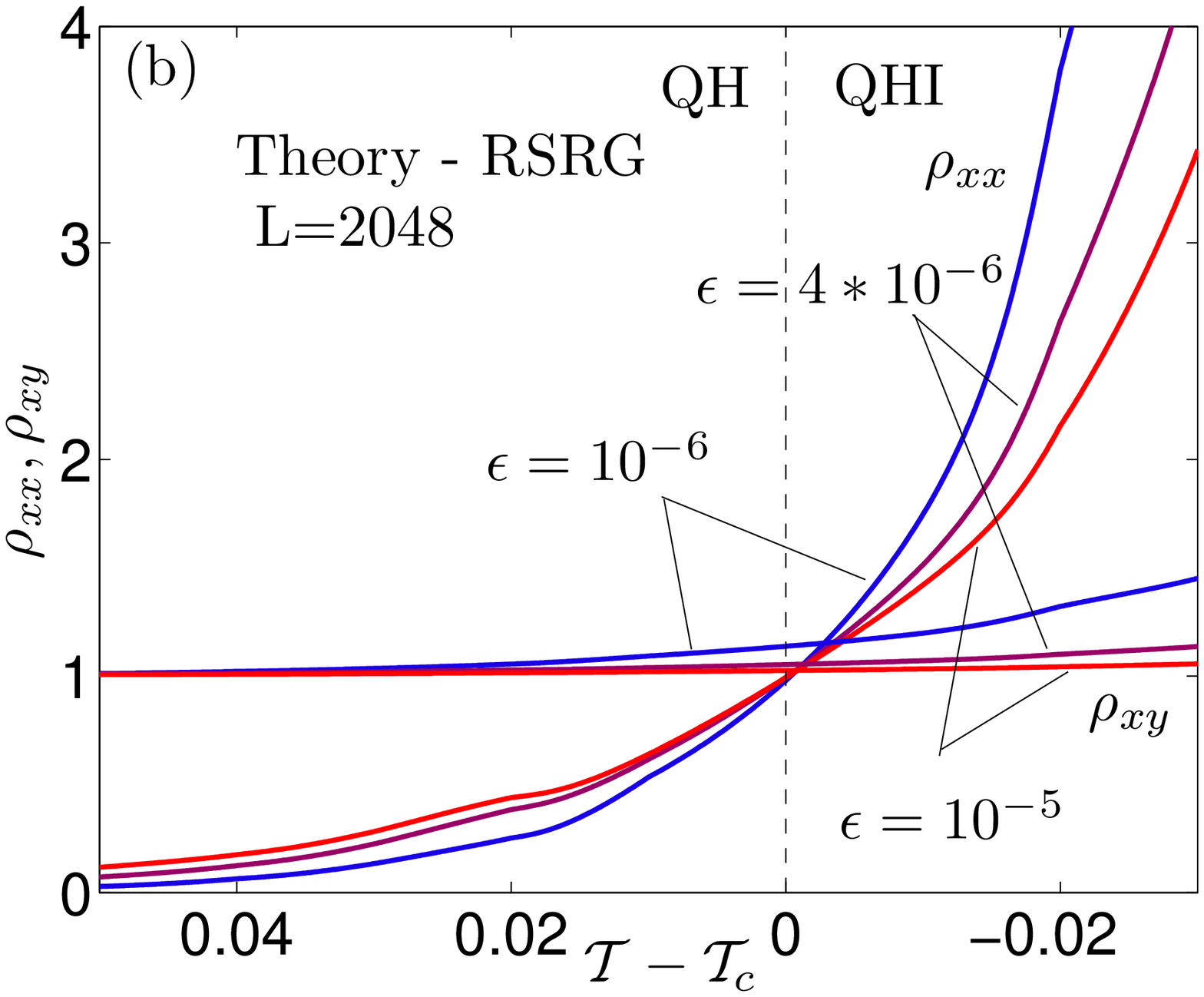}
\includegraphics[clip,width=0.45\hsize ]{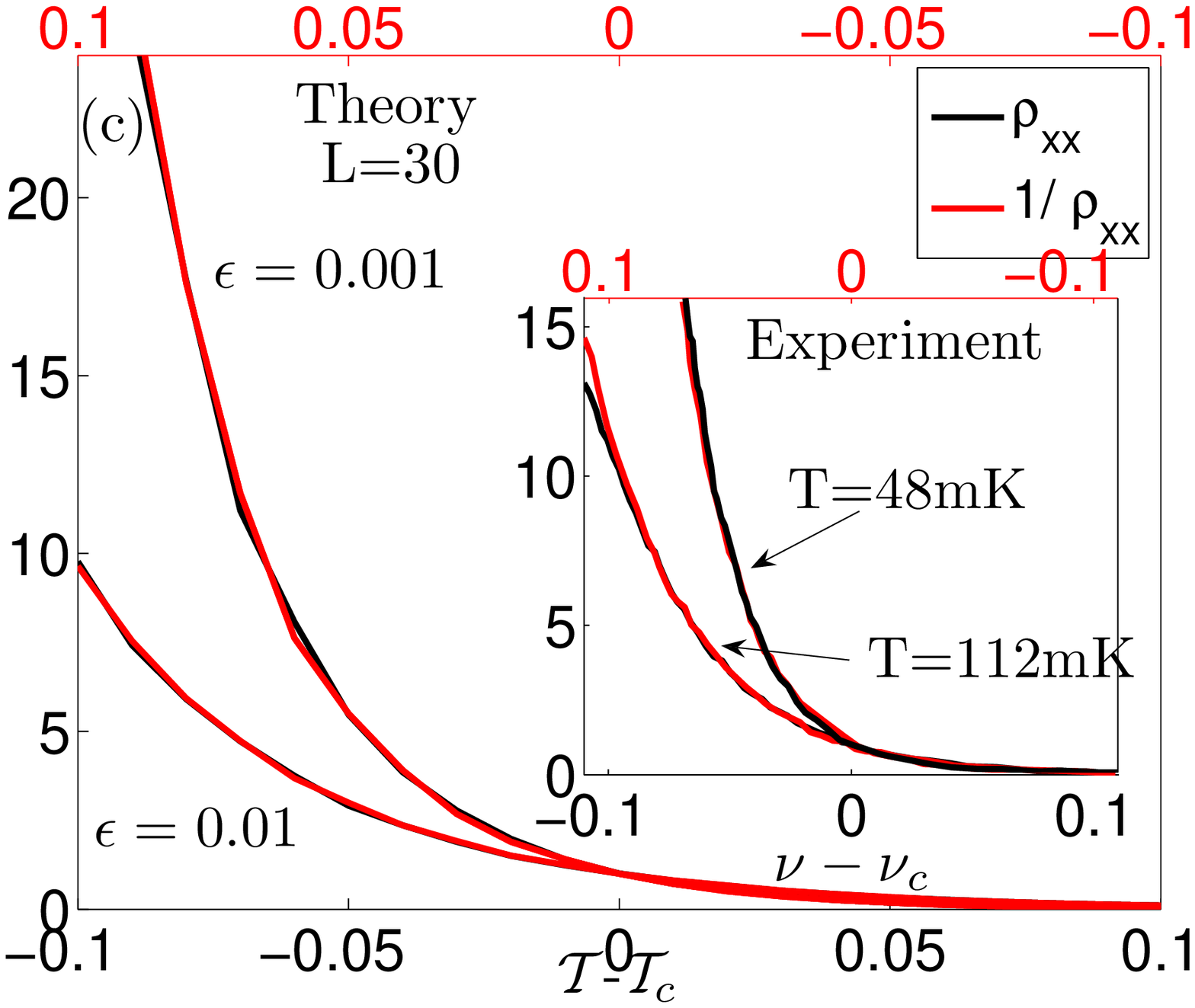}
\includegraphics[clip,width=0.45\hsize ]{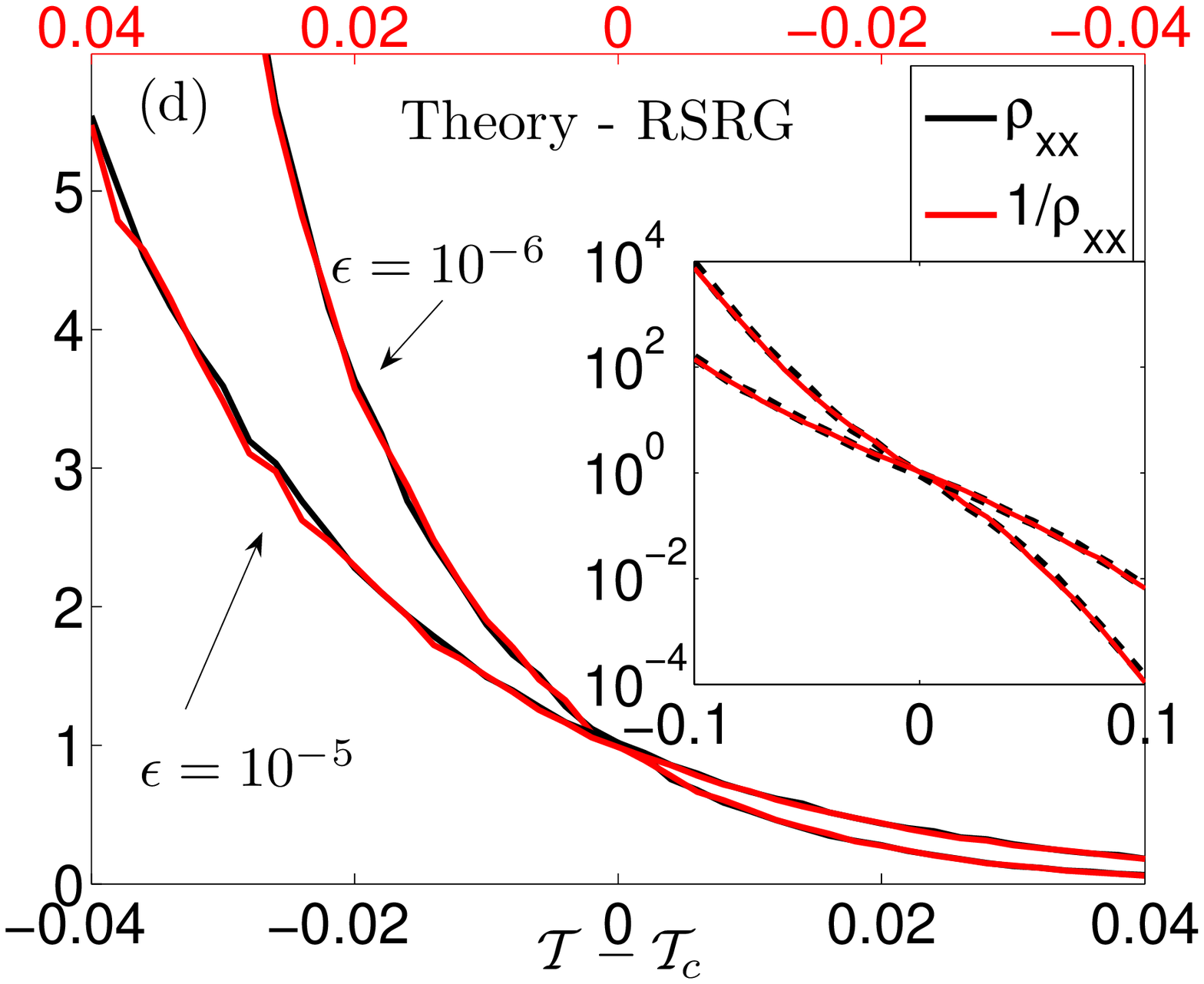}
\caption{(color online) Theoretical \textbf{(a,b)} and experimental \cite{Shahar1} (inset of \textbf{(a)}) results for the longitudinal and Hall resistivity plotted  as function of the deviation from the critical point for  systems with different phase lengths (theory) or temperatures  (experiments). The quantization of the Hall resistance on the insulating side improves with increasing decoherence (theory) or temperature (experiment). Symmetry between the quantum Hall phase and the insulating phase. $\rho_{xx}$ on the insulating side and $1/\rho_{xx}$ on the quantum Hall side, experiment \cite{Shahar1} (inset of \textbf{(c)}) and theory \textbf{(c,d)}. Theoretical results are for system with finite phase length. The data is plotted with the axis for $1/\rho_{xx}$ inverted (top axis),  for two different temperatures (experiment) and phase lengths (theory), demonstrating the symmetry between the two phases. The inset demonstrates that this symmetry is obeyed in the theory for many orders of magnitude.}
\label{fig:4}
\end{figure}
Having established the stability of the quantum Hall insulator phase, we compare our results, using both methods, in Fig.~\ref{fig:4}, to the experimental data. Panels (a) and (b) depict our calculation and the experimental data \cite{Shahar1} (inset of (a)). $\rho_{xx}$ and $\rho_{xy}$ are plotted as a function of the distance from the critical point ($\T-\T_c$ in the theoretical curves, $B-B_c$ in the experimental curve). Several theoretical curves, for a system of fixed $L$, but of different $\e$ (or $L_\phi$) are plotted, demonstrating that the quantization of $\rho_{xy}$ in the quantum Hall insulator phase becomes more exact as the level of decoherence increases (larger $\e$, smaller $L_\phi$). Interestingly, the experimental curves exhibit better quantization with increasing  temperature, which we attribute to increased decoherence. Enhanced decoherence also explains  the  better quantization of $\rho_{xy}$  for higher currents \cite{Hilke98}. For even higher temperatures, approaching the energy gap in the quantum Hall regime, one observes breakdown of the quantization in both phases - the quantum Hall phase and in the quantum Hall insulator phase \cite{Hilke98,Hilke99,Hilke99a,delang07}.
Another striking feature of the experimental data \cite{Shahar1} was the symmetry of $\rho_{xx}$ on the two sides of the critical point, $\rho_{xx}(\Delta\nu)=1/{\rho_{xx}(-\Delta\nu)}$,
where $\nu$ is the filling factor, the number of electrons in the system per available states in a Landau level, inset of Fig.~\ref{fig:4}c. This symmetry is also manifested in our results (Fig.~\ref{fig:4}c,d). In the coherent case it can be traced to the symmetry of the disorder potential, which leads to $\T_c=1-\T_c=1/2$. In that case it is easy to see \cite{cain} that since, by definition, in the fully coherent case $\rho_{xx}(\T)=\T/(1-\T)$, then clearly $\rho_{xx}(\T_c+\Delta\T) = (\T_c+\Delta\T)/(1-\T_c-\Delta\T) = 1/\rho_{xx}(\T_c-\Delta\T)$. In the presence of incoherent scattering, $\rho_{xx}$ is given, as discussed above, by the coherent $\rho_{xx}$ on a scale of $L_\phi$. Since in the model the decoherence parameter is, by definition (Eq.~\ref{1drg})  symmetric around the critical point, then the above relation is obeyed perfectly (see inset of Fig.~\ref{fig:4}d). The experimental deviations from this symmetry (inset of Fig.~\ref{fig:4}c) thus makes it possible to investigate the dependence of the decoherence length on magnetic field and density, allowing a deeper understanding of the nature of the incoherent processes at such low temperatures.

The relevance of incoherent scattering at milli-kelvin temperatures in the quantum Hall regime, imperative, as shown here, for explaining the quantum Hall insulator phase, has already been established experimentally \cite{Tsui}.  Incoherent scattering should be explored in the context of other quantum phase transition as well. In particular, it may also explain other puzzling two-dimensional phenomena, such as the apparent metal-insulator transition \cite{Kravchenko,Hanein}, or the intermediate
metallic phases observed in the superconductor-insulator transition in disordered thin films \cite{Mason}, and in the quantum Hall to insulator
transition \cite{Huang}. The present calculation allows quantitative  determination of the incoherence length which is crucial, for example, for any
possible application of mesoscopic devices as quantum bits, the basic building blocks of a quantum computer.

We thank A. Auerbach and A. Stern for fruitful discussions. This work was supported by the ISF and BSF.

%



\begin{thebibliography}{25}

\bibitem{vonklizing80}
K. v. Klitzing, G. Dorda and M. Pepper, Phys. Rev. Lett. \textbf{45}, 494 (1980).

\bibitem{Shahar1}
D. Shahar \textit{et al}., Solid State Commun. \textbf{102}, 817-821 (1997).

\bibitem{Shahar2}
D. Shahar \textit{et al}.,  Phys. Rev. Lett. \textbf{79}, 479-482 (1997).

\bibitem{Hilke98}
M. Hilke \textit{et al}., Nature \textbf{395}, 675-677 (1998).

\bibitem{Hilke99}
M. Hilke \textit{et al}., Ann. Phys. \textbf{8}, 603 (1999).

\bibitem{Hilke99a}
M. Hilke \textit{et al}., Europhys. Lett. \textbf{46}, 775 (1999).

\bibitem{visser06}
A. de Visser \textit{et al}., J. Phys.: Conf. Ser. \textbf{51}, 379 (2006).

\bibitem{delang07}
D. T. N. de Lang  \textit{et al}., Phys. Rev. \textbf{B 75},  035313 (2007).


\bibitem{Entin-Wohlman}
O. Entin-Wohlman \textit{et al}., Phys. Rev. Lett. \textbf{75}, 4094 (1995).
\bibitem{Pryadko}
L. P. Pryadko and A. Auerbach, Phys. Rev. Lett. \textbf{82}, 1253 (1999).

\bibitem{Zulicke}
U. Z\"ulicke and E. Shimshoni, Phys. Rev. B \textbf{63}, 241301(R)  (2001).

\bibitem{cain}
P. Cain and R. A. R\"omer, Europhys. Lett. \textbf{66}, 104 (2004).

\bibitem{Fertig87}
H. A. Fertig and B. I. Halperin, Phys. Rev. B \textbf{36}, 7969 (1987).

\bibitem{Trugman}
S. A. Trugman, Phys. Rev. B \textbf{27}, 7539  (1983).

\bibitem{Milnikov}
G. V. Mil'nikov and I. M. Sokolov, JETP Lett. \textbf{48}, 536 (1988).

\bibitem{Dubi}
Y. Dubi, Y. Meir and Y. Avishai, Phys. Rev. B \textbf{71}, 125311 (2005).

\bibitem{Chalker}
J. T. Chalker and P. D. Coddington, J. Phys. C \textbf{21}, 2665 (1988).


\bibitem{Galstyan97}
A. G. Galstyan and M. E. Raikh, Phys. Rev. B \textbf{56}, 1422 (1997).

\bibitem{Huckenstein}
B. Huckenstein, Rev. Mod. Phys. \textbf{67}, 357 (1995).

\bibitem{Pruisken}
A. M. M. Pruisken, Phys. Rev. B \textbf{32}, 2636 (1985).

\bibitem{sheng}
D. N. Sheng and Z. Y. Weng, Phys. Rev. B \textbf{59}, R7821 (1999).

\bibitem{buttiker}
 M. B\"uttiker, Phys. Rev. B \textbf{33}, 3020 (1986).

\bibitem{Landauer-Buttiker}
M. B\"uttiker, \textit{et al.}, Phys. Rev. B \textbf{31}, 6207 (1985).

\bibitem{Shimshoni97}
E. Shimshoni and A. Auerbach,  Phys. Rev. B  \textbf{55}, 9817 (1997).

\bibitem{Dikhne}
A. M. Dykhne and I. M. Ruzin, Phys. Rev. B \textbf{50}, 2369 (1994).

\bibitem{Tsui}
Wanli Li \textit{et al.},  Phys. Rev. Lett. \textbf{102}, 216801  (2009).

\bibitem{Kravchenko}
 S. V Kravchenko, \textit{et al.},  Phys. Rev. B \textbf{50}, 8039  (1994).

\bibitem{Hanein}
Y. Hanein, \textit{et al.}, Phys. Rev. B \textbf{58}, R13338 (1998).


\bibitem{Mason}
N. Mason and A. Kapitulnik, Phys. Rev. B \textbf{64}, 060504 (2001).

\bibitem{Huang}
T-Y Huang,  \textit{et al.}, Phys. Rev. B \textbf{78}, 113305 (2008).




\bibitem{MC}
P. Cain, R. A. R\"omer, M. Schreiber and M. E. Raikh, Phys. Rev. B \textbf{64}, 235326 (2001).


\end{thebibliography}
\end{document}